\documentclass[twocolumn,showpacs,preprintnumbers]{revtex4}
\usepackage{amssymb}
\usepackage{amsfonts}
\usepackage{amsmath}
\usepackage{graphicx}
\usepackage{dcolumn}
\usepackage{bm}

\begin{document}

\title{Microwave induced magnetoresistance oscillations at the
 subharmonics of the cyclotron resonance.}
\author{S.~I.~Dorozhkin,$^{1,2}$ J.~H.~Smet,$^1$ K.~von Klitzing$^1$, L.~N.~Pfeiffer,$^3$ and K.~W.~West$^3$}
\affiliation{$^1$Max-Planck-Institut f\"{u}r
Festk\"{o}rperforschung,
 Heisenbergstra\ss e 1, D-70569 Stuttgart, Germany}
\affiliation{$^2$Institute of Solid State Physics, Chernogolovka,
Moscow district, 142432, Russia} \affiliation{$^3$Bell
Laboratories, Lucent Technologies, Murray Hill, New Jersey 07974 }

\begin{abstract}
The magnetoresistance oscillations, which occur in a
two-dimensional electron system exposed to strong microwave
radiation when the microwave frequency $\omega$ coincides with the
\emph{$n$-th} subharmonic of the cyclotron frequency $\omega_{\rm
c}$ have been investigated for $n= 2,3$ and $4$. It is shown that
these subharmonic features can be explained within a
non-equilibrium energy distribution function picture without
invoking multi-photon absorption processes. The existence of a
frequency threshold above which such oscillations disappear lends
further support to
 this explanation.
\end{abstract}

\date{\today}
\maketitle

The observation of microwave induced magnetoresistance
oscillations (MIMO) with some minima saturating near zero in
state-of-the-art two-dimensional electron
systems~\cite{zudov0,mani,zudov1} has attracted great interest.
The effect is governed by the ratio of the microwave frequency to
the cyclotron frequency: $\omega/\omega_{\rm c}$. At the magnetic
field close to an integer ratio, the magnetoresistance remains
unaffected by the microwave radiation (zero-response node). At
slightly larger values of $\omega/\omega_{\rm c}$, the microwaves
cause a drop of resistance, while at smaller values they produce a
positive contribution to the resistance. The two mainstream
approaches to explain these oscillations are based on scattering
assisted indirect optical transitions~\cite{ryzhii1,durst} and on
the creation of a non-equilibrium electron energy distribution
function~\cite{dorozh1,dmitriev1}. Already in early
experiments~\cite{zudov1,dorozh1,zudov2}, additional oscillations
were detected near $\omega=\omega_{\rm c}/2$ as well as close to
other fractional values of $\omega/\omega_{\rm c}$ such as 3/2,
5/2, and 2/3. In a recent article~\cite{zudov3}, these
'fractional' oscillations were tentatively ascribed to
multi-photon processes involving the simultaneous absorption of
$n$ photons. Here, $n = 2$ or $3$ and corresponds to the
denominator of the fractional value. This interpretation was
supported on a qualitative level within the model of microwave
induced indirect transitions when multi-photon absorption is
included~\cite{lei1,apel,lei2}.

In this paper, we report on the observation of MIMO when
$\omega=\omega_{\rm c}/n$ with $n=2,3$ and $4$ and in particular
we investigate the frequency range where these additional
oscillations are observed. We find that they only occur at
microwave frequencies below a certain threshold value. If
interpreted in terms of indirect optical transitions our data
would imply the involvement of two-, three- and four-photon
absorption processes. However within this framework, it is
difficult to account for the frequency threshold. We show that all
important properties of these additional oscillations, including
the frequency threshold, are captured by the non-equilibrium
electron distribution function picture when only single photon
absorption processes are considered.

Our investigations were carried on a double-sided modulation doped
GaAs/AlGaAs quantum well with an electron density of $n_{\rm
s}=2.7\times 10^{11}{\rm cm}^{-2}$ and a mobility $\mu=17 \times
10^{6}\ {\rm cm}^2/{\rm V\,s}$. Hall bar geometries were prepared
with a width $W$ of $400\ {\rm \mu m}$ and with potential probes
spaced a distance $L = 800\ {\rm \mu m}$ apart. The sample was
placed near the end of a short circuited microwave waveguide with
a cross-section of $16\times 8\ {\rm mm}^2$ (WG18). Waveguide and
sample were submerged in pumped $^3{\rm He}$. To control the
microwave power incident on the sample, a carbon resistor was
mounted close to the sample. The dissipative ($\rho_{xx}$) and
Hall ($\rho_{xy}$) resistivity components were measured at 10 Hz
with a lock-in technique.

Fig.~1 illustrates how the MIMO near $\omega=\omega_{\rm c}/n$
arise with increasing microwave power. At small power levels, only
the MIMO at the harmonics of $\omega_{\rm c}$ (when $\omega= n
\omega_{\rm c}$) are observed at weak magnetic fields. For
example, the curve for $P=-18\ {\rm dBm}$ reveals two oscillations
located near $\omega=\omega_{\rm c}$ and $\omega=2\omega_{\rm c}$.
With increasing power, oscillations emerge at subharmonics of the
cyclotron frequency, i.e.~when $\omega = \omega_{\rm c}/n$. At the
highest incident power of $9\ {\rm dBm}$, the $n=4$ subharmonic
can be discerned. These oscillations appear for both signs of the
magnetic field. The amplitudes of all MIMO do not change
monotonically with power, but pass through a maximum.
\begin{figure}[tb]
\includegraphics{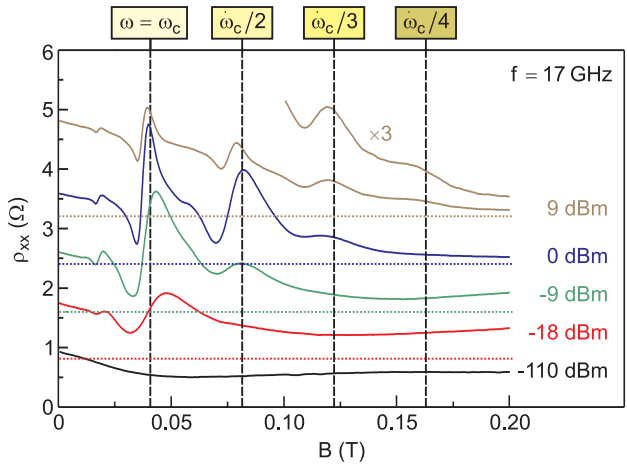}
\caption{Magnetoresistivity $\rho_{\rm xx}$ versus $B$ for
different microwave power levels. The microwave frequency
$f=\omega/2\pi$ is equal to 17 GHz. For clarity, the curves are
offset vertically. The horizontal dotted lines mark zero
resistance. Vertical dashed lines indicate the position of the CR
and its subharmonics calculated for the electron effective mass of
$0.067\,m_{\rm e}$. To bring out better the microwave induced
oscillation at the 4th subharmonic, part of the upper curve has
been magnified.} \label{Fig1}
\end{figure}

Fig.~2 demonstrates the existence of zero-response nodes near the
location of the cyclotron resonance and its second subharmonic.
There, the magnetoresistance does not depend on the microwave
power and hence curves recorded at different power levels
intersect (for MIMO at $\omega = n \omega_{\rm c}$ these nodes
were considered in some detail in Refs.~\cite{mani,mani1}).
Similar nodes do not seem to exist at higher order subharmonics
($n=3,4$) and the photoresistivity lies entirely underneath the
dark curve ($P=-110\ {\rm dBm}$-trace) at least for the
experimentally attainable amplitudes of the oscillations. Note
that both the zero response nodes and the oscillations associated
with the 3d and 4-th subharmonic are shifted to a slightly lower
$B$-field than calculated from $\omega = \omega_{\rm c}/n$
\emph{($n=1,2,3,4$)} when using an electron effective mass of
$0.067m_{\rm e}$ for GaAs. The minimum and maximum of each
oscillation come closer to the node at larger microwave power.
This has also been observed previously for MIMO near $\omega=n
\omega_{\rm c}$ (for example in Ref.~\onlinecite{dorozh1}) and can
be seen in Fig.~1 for the oscillation located at the cyclotron
resonance condition. At the highest power the minimum located near
$\omega\approx 2\omega_{\rm c}/3$ becomes visible and has been
marked by an arrow in Fig.~2.

A crucial property of the microwave induced oscillations at
subharmonics of the cyclotron resonance is that they are only
observed below a certain threshold frequency. In Fig.~3 some
traces taken at different frequencies are compared. In these
experiments, the microwave power at the oscillator output was
adjusted at each frequency to ensure the same signal on the
bolometer located close to the sample. The oscillation near
$\omega = \omega_{\rm c}/2$ appears rather abruptly at microwave
frequencies below 40 GHz. The oscillation is absent at 40 GHz,
weakly present at 35 GHz and has drastically increased in
amplitude at 30 GHz. The amplitudes of the microwave induced
oscillations at harmonics of $\omega_{\rm c}$ behave oppositely.
In this frequency regime, they come out stronger at higher
frequencies. Note that also in Ref.~\cite{zudov1}, the second
subharmonic feature was reported to be visible only below 50 GHz.

\begin{figure}[tb]
\includegraphics[width=8.5cm,clip]{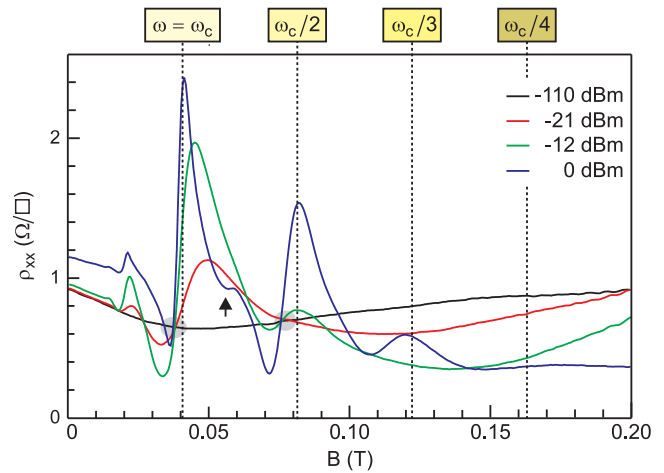}
\caption{The magnetoresistivity $\rho_{\rm xx}$ versus $B$ as in
Fig.~1 for different microwave power levels but without vertical
offsets. The curves intersect at the center of the grey dots.}
\label{Fig2}
\end{figure}

\begin{figure}[tb]
\includegraphics{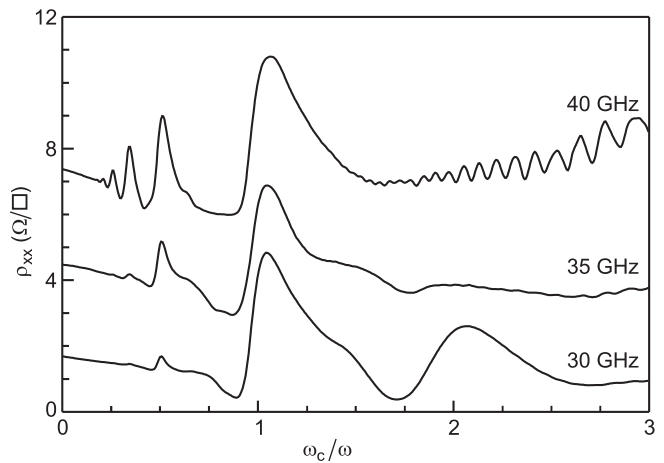}
\caption{The magnetoresistivity $\rho_{\rm xx}$ versus
$\omega_{\rm c}/\omega$ for three different microwave frequencies.
The data were measured at $0.5\ {\rm K}$. Curves were offset
vertically for clarity. The microwave power at the oscillator
output was adjusted to maintain a constant bolometer signal: $P(30
{\rm GHz})=2.2 {\rm mW}$, $P(35 {\rm GHz})=3.5 {\rm mW}$, and
$P(40 {\rm GHz})=5.0 {\rm mW}$.} \label{Fig3}
\end{figure}

We will show below in detail that single photon processes are
capable of accounting for oscillations at magnetic fields where
$\omega$ coincides with a subharmonic of $\omega_{\rm c}$. First
however we discuss the origin of the threshold frequency.
Qualitatively MIMO appear within the non-equilibrium occupation
picture as a result of transitions between two different broadened
Landau levels~\cite{dorozh1}. Such interlevel transitions are
possible provided that $\omega>\omega_{\rm c}-2\Gamma/\hbar$ with
$\Gamma$ being the half-width of a Landau level (Throughout this
manuscript, spin-splitting is neglected in view of the low
magnetic fields). For a $B$-field at which $\omega = \omega_{\rm
c}/n$, this inequality reduces to $\omega_{\rm
c}(1-1/n)<2\Gamma/\hbar$. This condition is fulfilled only below a
certain, sample and $n$-dependent threshold magnetic field $B_{\rm
th}^{n}$ because of the disparate functional dependencies of
$\omega_{\rm c}$ and $\Gamma$ on $B$: $\omega_{\rm c}$ is
proportional to $B$ whereas $\Gamma$ exhibits a weaker dependence
(for example, in the self-consistent Born approximation $\Gamma
\propto \sqrt{B}$~\cite{ando,dmitriev0} for non-overlapping Landau
levels, i.e.~when $\hbar\omega_{\rm c}>2\Gamma$). Above this
threshold magnetic field $B_{\rm th}^{n}$, the inter-Landau level
transitions at the position of the $n$-th subharmonic are not
possible and no oscillation arises. The corresponding frequency
threshold $f_{\rm th}^{n}$ equals $\omega_{\rm c}(B_{\rm
th}^n)/2\pi n$. It drops with increasing $n$. For the special case
when $n = 2$, the inter-Landau level transitions disappear at the
same magnetic field as intra-level transitions do. Hence, at
microwave frequencies above $f^{2}_{\rm th}$ a photoresponse is
absent all together for moderate microwave power levels in a
magnetic field interval around $\omega = \omega_{\rm c}/2$. This
has been demonstrated previously in Ref.~\onlinecite{dorozh2}. In
the sample investigated here, this 'zero response'-region appears
at a frequency of approximately 40 GHz and indeed the MIMO at the
second subharmonic develops only at lower frequencies (see
Fig.~3).

From this magnetic field interval where the magnetoresistance does
not respond to microwave radiation, it is possible to estimate the
homogeneous broadening $\Gamma$ of the Landau levels at these
$B$-fields. Assuming a square-root magnetic field dependence of
$\Gamma$, levels no longer overlap at $B > 0.045\ {\rm T}$. This
value is also consistent with the field $B\approx 0.05 {\rm T}$
above which Shubnikov - de Haas oscillations become pronounced in
this sample at low temperature ($< 100\ {\rm mK}$). This crossover
from overlapping to separated levels determined from the Shubnikov
- de Haas oscillations appears at a somewhat higher magnetic field
due to inhomogeneous level broadening, which smears the
oscillations but does not affect microwave-induced transitions
between levels. We conclude that the features at the subharmonics
of $\omega_{\rm c}$ in Fig.~1 and 2 exist in the regime of
separated Landau levels. Hence, we will consider here theoretical
calculations, which are based on a non-equilibrium electron
distribution function and assume separated levels.

For the DC diagonal magnetoconductivity $\sigma_{\rm xx}^{\rm dc}$
of the 2DES subjected to microwave radiation we adopt the
following formula from Refs.~\onlinecite{dorozh1}
and~\onlinecite{dmitriev1}:
\begin{equation}
\sigma_{{\rm xx}}^{{\rm dc}}=\int \sigma(\epsilon)
\left(-\frac{\partial f}{\partial\epsilon}\right) d\epsilon.
\end{equation}
\noindent Here $\sigma(\epsilon)$ equals $\frac{n_{\rm
s}e^2\tau}{m^*(\omega_{\rm c}\tau)^2}\tilde{\nu}^2(\epsilon)$,
$\tilde{\nu}(\epsilon)=\nu(\epsilon)m^*/\pi\hbar^2$ is the density
of states in a quantizing magnetic field normalized to its
zero-field value, $\tau$ is the momentum relaxation time and
$f(\epsilon)$ is a steady state non-equilibrium distribution
function. The solution of the kinetic equation within the energy
relaxation time approximation yields the following recursive
formula for $f(\epsilon)$~\cite{dmitriev1}:
\begin{equation}
P_{\rm
\omega}\sum_{\pm}\tilde{\nu}(\epsilon\pm\hbar\omega)[f(\epsilon\pm
\hbar\omega)-f(\epsilon)]=f(\epsilon)-f_{\rm T}(\epsilon),
\end{equation}
\noindent where $f_{\rm T}$ is the Fermi distribution function.
The dimensionless quantity $P_{\omega}$ is a measure for the
incident microwave power and is given by
\begin{equation}
P_{\omega}=\frac{\tau_{\rm in}}{4\tau}\left(\frac{eE_{\rm \omega}
v_{\rm F}}{\hbar\omega}\right)^2\frac{\omega_{\rm
c}^2+\omega^2}{(\omega^2-\omega_{\rm c}^2)^2}.
\end{equation}
\noindent Here $\tau_{\rm in}$, is the energy relaxation time of
photo-excited electrons~\cite{comment2}, $E_{\rm \omega}$ is the
amplitude of the microwave electric field, and $v_{\rm F}$ is the
Fermi velocity. The dc-resistivity $\rho_{\rm xx}$ equals
$\sigma_{\rm xx}/\sigma_{\rm xy}^2$ with $\sigma_{\rm xy}= n_{\rm
s}e/B$. The non-equilibrium distribution function theory in
Ref.~\onlinecite{dmitriev1} has been developed within the
self-consistent Born approximation and the density of states of a
separated Landau level ($\hbar\omega_{\rm c}>2\Gamma$) takes on a
semi-elliptical form~\cite{comment3}:
\begin{equation}
\nu(\epsilon)=(2N_0/\pi\Gamma)\sum_{n=0}Re\,\left[1-\left(\frac{\epsilon-(n+1/2)\hbar\omega_{\rm
c}}{\Gamma}\right)^2\right]^{1/2}.
\end{equation}
\noindent Here, $N_0=eB/h$ denotes the Landau level degeneracy and
the half-width of a disorder-broadened Landau level is determined
from $\Gamma=\hbar\sqrt{2\omega_{\rm c}/\pi\tau_{\rm q}}$ with
$\tau_{\rm q}$ being the quantum scattering time.
\begin{figure}[tb]
\includegraphics[width=8.5cm,clip]{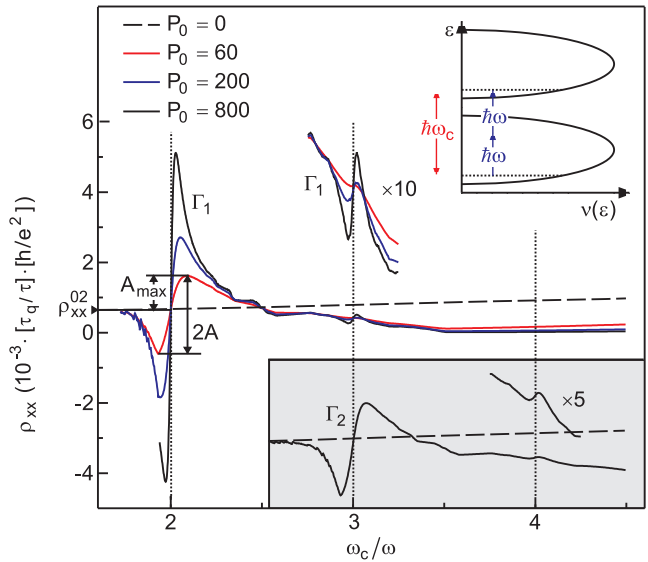}
\caption{Calculated magnetoresistivity versus $\omega_{\rm
c}/\omega$ for different power levels $P_{\rm 0} \equiv
P_{\omega}(\omega_{\rm c }=0)$. The calculations were performed
with the following parameter values: $\hbar\omega\approx 6.4\times
10^{-3}\epsilon_F^0$, $T \approx 1.2\times 10^{-2}\epsilon_F^0$,
and $\Gamma_1\approx \hbar\sqrt {0.45\omega\omega_{\rm c}}$. Here
$\epsilon_F^0$ is the Fermi energy at $B = 0$, $T$ is the
temperature, and $\Gamma_{1}$ is the width of the Landau levels.
The data shown in the inset have been calculated for $\Gamma_2
\approx \hbar\sqrt{0.7\omega\omega_{\rm c}}$ and $P_{0}=800$. The
top inset illustrates for the case of the 2-nd subharmonic that
both intra- and inter-level transitions play a role.} \label{Fig4}
\end{figure}

Numerical calculations based on Eqs.~(1)-(4) are depicted in
Fig.~4. They demonstrate the existence of MIMO at the second,
third (main panel) and fourth subharmonic (the inset at the
bottom) of the cyclotron frequency~\cite{comment}. These curves
confirm the following experimental features: (i) the shape of the
MIMO observed in the experiment, (ii) the existence of a
zero-response node near $\omega = \omega_{\rm c}/2$ where the
magnetoresistance is insensitive to the radiation, (iii) the
appearance of 3d and 4th subharmonic features below the dark
curve, and (iv) the large microwave power levels required to make
features at the higher order subharmonics appear.

Even though these subharmonic features reflect a commensurability
between the cyclotron and the microwave frequencies, this
commensurability effect is not as trivial as in the case of the
features at multiples of $\omega_{\rm c}$. These subharmonic
features require that the microwave frequency, the width of the
Landau levels and the distance between adjacent Landau levels are
such that the microwave radiation can trigger both inter- and
intra-Landau level transitions as illustrated schematically in the
inset of Fig.~4. This combination of a microwave induced
redistribution of electrons within the Landau level with
inter-level transitions establishes a connection between states of
neighboring levels separated by an energy $\hbar \omega_{\rm c}$
even though the microwave photon energy is only a fraction of the
cyclotron energy. We stress that there is no need for the
simultaneous absorption of multiple photons. On the contrary, if
multi-photon absorption processes are included, the theory would
not predict a frequency threshold as is observed in experiment.
The distribution function scenario with one-photon absorption
processes only can also account for microwave induced features at
fractional ratios of $\omega/\omega_{\rm c}$ with a nominator
different from unity~\cite{dorozh3}.

\begin{figure}[tb]
\includegraphics[width=8.5cm,clip]{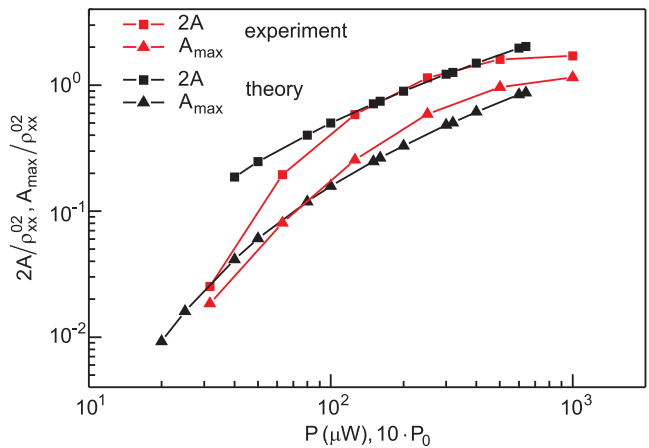}
\caption{Measured (red signs) and calculated (black signs) values
of the maximum $A_{\rm max}$ (triangles) and the total amplitude
$2A$ (squares) of the magnetoresistivity oscillation induced by
the microwave radiation when $\omega = \omega_{\rm c}/2$ versus
the microwave power at the oscillator output $P$ or the
dimensionless parameter $P_0$. The amplitudes are normalized to
$\rho_{\rm xx}^{\rm node}$ (marked in Fig.~4). The experimental
and theoretical power scales are adjusted to get the best
agreement between calculated and measured data for the total
oscillation amplitude. The parameters of the calculations are the
same as for the main panel of Fig.~4.} \label{Fig5}
\end{figure}

Additional support for our interpretation comes from a power
dependent study of two quantities $2A$ and $A_{\rm max}$ defined
in Fig.~4. These quantities characterize the amplitude of the
oscillation at the 2nd subharmonic. A comparison of the
experimental and theoretical dependencies is presented in Fig.5.
Plotted values have been normalized to the magnetoresistance value
$\rho_{\rm xx}^{\rm node}$ at the zero response node (marked by a
triangle on the ordinate of Fig.~4). The only fitting parameter in
Fig.~5 is a constant coefficient relating the microwave power
measured at the oscillator output to the power $P_0$ used in the
calculations. This fit parameter does not affect the slopes of the
curves, but merely shifts curves on the logarithmic horizontal
axis. We note that it is very difficult to determine the absolute
value of the microwave power incident on the sample, however at
the same frequency this incident power is proportional to the
power emitted by the oscillator. Reasonable agreement is obtained
between experiment and theory with this single parameter. In
particular, the decreasing slope with increasing power is
reproduced by the theory. Furthermore, theoretical and
experimental slopes agree well when $2A/\rho_{\rm xx}^{\rm node}$
is close to unity.

In conclusion, microwave induced oscillations which emerge at
large incident power when the microwave frequency coincides with a
subharmonic of the cyclotron frequency have been experimentally
studied and accounted for successfully within the non-equilibrium
distribution function picture. This model does not rely on the
simultaneous absorption of multiple photons. The existence of a
frequency threshold above which these oscillations are absent
further corroborates this picture.

We acknowledge financial support from INTAS, RFBR (SID)and the
DFG. We also thank I.~Dmitriev for fruitful comments on the
manuscript.



\begin{thebibliography}{99}
\bibitem{comment2} Here we use a simplified model which ignores
a radiation induced change of the electron temperature. Such a
change affects the Fermi distribution function $f_{\rm T}$ (Eq.2)
and the  energy relaxation time $\tau_{\rm in}$
(Eq.3)~\cite{dmitriev1}. Taking into account such electron heating
effects, however only modifies the amplitude of the MIMO.
\bibitem{comment3} Strictly speaking, the density of states
takes on the semi-elliptical form when $\hbar\omega_{\rm
c}>>2\Gamma$, which is not fullfilled in our case in view of the
constraint $\omega_{\rm c}(1-1/n)<2\Gamma/\hbar$. However,
deviations from the semi-elliptical density of states are already
small in this crossover regime, $\hbar\omega_{\rm c}\sim 2\Gamma$,
provided that the Landau levels are separated ($\hbar\omega_{\rm
c}>2\Gamma$)~\cite{dorozh3}.
\bibitem{comment} For the square-root magnetic field dependence of
the Landau-level width assumed in our calculations, it is not
possible that inter-Landau level transitions are allowed at the
4th subharmonic while simulaneously having separated Landau levels
at magnetic fields corresponding to the 2nd subharmonic. Hence
under the assumption of separated Landau levels for one and the
same choice of parameter $\Gamma$, it is not possible to
simultaneously obtain MIMO at these two subharmonics.
\end{thebibliography}
\end{document}